\documentclass[fleqn,12pt]{wlscirep}
\usepackage[utf8]{inputenc}
\usepackage{authblk}
\usepackage{hyperref}
\usepackage{graphicx}
\usepackage{amsmath}
\usepackage{amssymb}
\usepackage[T1]{fontenc}
\usepackage{bm}
\usepackage{tabularx}[xcolor]
\usepackage{booktabs,array,ragged2e}
\usepackage{makecell,xstring,collcell} 
\usepackage{threeparttable}
\usepackage{etoolbox}
\usepackage{microtype}
\usepackage{mathtools}
\usepackage{xcolor}

\newcolumntype{Y}{>{\RaggedRight\arraybackslash}X}
\newcolumntype{P}[1]{>{\RaggedRight\arraybackslash}p{#1}}

\newcommand{\xbreak}[1][.5ex]{\par\vspace{#1}}

\usepackage{setspace}            
\title{Radioactive Molecules as Laboratories of Fundamental Physics}
\author[1,2]{A. Jadbabaie}
\author[1,2]{S. Ebadi}
\author[1]{R.F. Garcia Ruiz}
\author[3]{N. R. Hutzler}
\author[4]{A. M. Jayich}
\author[5]{J. T. Singh}
\affil[1]{Department of Physics, Massachusetts Institute of Technology, Cambridge, MA}
\affil[2]{Department of Physics, Harvard University, Cambridge, MA, USA}
\affil[3]{Division of Physics, Mathematics, and Astronomy, California Institute of Technology, Pasadena, CA, USA}
\affil[4]{Department of Physics, University of California, Santa Barbara, CA, USA}
\affil[5]{Facility for Rare Isotope Beams, Michigan State University, East Lansing, MI, USA}

\begin{abstract}
Radioactive molecules provide a powerful new platform {in the search for new} physics at energy scales complementary to high-energy particle colliders. By combining enhancements from nuclear properties with the {sensitivity and} control offered by molecular structure, experiments with radioactive molecules offer great {reach in the search for} new physics beyond the Standard Model. Rapid progress in this field is being driven by advances in the production and control of radioactive molecules, alongside the development of new experimental tools and theoretical techniques. In this Perspective, we discuss the current status and future prospects of this rapidly developing, interdisciplinary field at the intersection of nuclear physics, atomic and molecular physics, and {particle} physics.
\end{abstract}

\begin{document}
\maketitle

\section{Introduction}
\label{sec:intro}
Discovering new physics beyond the Standard Model, our {best} framework for describing elementary particles and their interactions, is a central goal in modern physics. Despite its remarkable success, the Standard Model cannot explain open questions about the origin and evolution of the universe. 
One of the most pressing mysteries is the observed imbalance between matter and antimatter. According to the known laws of physics, the Big Bang should have produced equal amounts of both; yet, observations reveal that the visible universe consists almost entirely of matter, with antimatter contributing at most one part per billion{~\cite{CanettiMatter2012}}.
Possible mechanisms involve processes, governed by yet-to-be discovered particles in the early universe, that break fundamental{,} discrete symmetries of nature, and hence {produce} more matter than {antimatter} \cite{Sakharov1967,DeVriesElectroweak2018}. 
This has motivated experimental and theoretical efforts across disciplines {to characterize the laws of physics under fundamental symmetries of charge-conjugation (C), parity-reversal (P), and time-reversal (T). }
Experiments are ongoing at the {energy frontier}, where colliders such as the Large Hadron Collider (LHC) directly probe new particles and forces \cite{Aaij2017, Bar-ShalomTheoretical2025}, {the intensity frontier with studies of neutrinos and rare processes~\cite{Artuso2022}}, and at the precision frontier, where low-energy measurements provide a powerful and complementary approach~\cite{Safronova2018,Chupp2019}. While the LHC is sensitive to new physics at the TeV scale, precision measurements can indirectly probe much higher energy scales by measuring {minute physical effects associated with the presence of undiscovered, symmetry-violating} particles or forces~\cite{Safronova2018}. 

{Extensions beyond the Standard Model generally introduce interactions that break fundamental symmetries, unless symmetry conservation is explicitly imposed~\cite{IsidoriMinimal2012,GreljoAdding2022}. Once a symmetry is broken at high energies,} { its effects can be traced through a ``tower'' of effective field theories (EFTs)~\cite{GeorgiEffective1993,ManoharIntroduction2020,HammerNuclear2020,Isidoristandard2024} to low energies, manifesting as symmetry-violating electromagnetic properties of nucleons, atoms, and molecules~\cite{Pospelov05, EngelElectric2013, Chupp2019}. The most prominent example of such a property is a permanent electric dipole moment (EDM) of a fundamental particle~\cite{AlarconElectric2022}, which is aligned with the particle's intrinsic spin by the Wigner-Eckart theorem. The correlation of spin pseudovector and electric dipole vector violates P and T symmetries, which by the CPT theorem implies violation of Charge-Parity (CP) symmetry. Within the Standard Model, only the weak force is known to violate C, P, and CP symmetries individually~\cite{Chupp2019}. However, EDMs induced by the weak force remain unobserved, as Standard Model contributions are at least six orders of magnitude below the current experimental sensitivity~\cite{SengReexamination2015,Chupp2019, DragosConfirming2021, EmaStandard2022}.}

{CP-violation (CPV) can also, in principle, occur in the strong force, via a topological term in the Standard Model Lagrangian~\cite{HookTASI2018} $\mathcal{L}_{\bar\theta}\propto \bar \theta$. Since $\mathcal{L}_{\bar\theta}$ is non-perturbative, its effects vanish for high-energy colliders, and $\bar \theta$ must be probed by low-energy physics, such as EDMs. To date, experiments have found no evidence for strong CPV, and in particular, bounds on the neutron EDM~\cite{Abel2020} have constrained the value of $\bar{\theta}< 10^{-10}$. Known as the strong CP problem, this striking observation of an unnaturally small unitless parameter, even after including radiative corrections~\cite{deVriesIndirect2019}, is a major puzzle of modern physics, and an important motivation for QCD axion models~\cite{Pec77,KimAxions2010,deVriesUncovering2021,AdamsAxion2023}. }

\begin{figure}[t]
    \centering
    \includegraphics[width=0.75\linewidth]{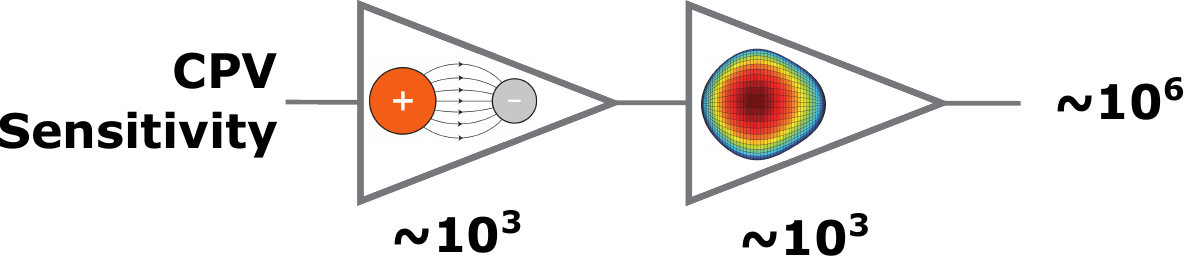}
    \caption{By combining the molecular enhancement {(left)} due to the extreme electromagnetic environment inside polar molecules, with the nuclear enhancement {(right)} offered by heavy, octupole-deformed nuclei, we can realize significant experimental enhancements to intrinsic CPV sensitivity. {Octupole-deformed nucleus image reproduced from Ref.~\cite{Gaffney2013}.}}
    \label{fig:amplifier}
\end{figure}

{Within atoms and molecules, the presence of CPV, i.e., an EDM, leads to small energy shifts that are amplified in heavy species due to interactions between finite-size nuclei and relativistic, core-penetrating electrons.} { In particular, polar molecules, which break spherical symmetry and are highly polarizable, exhibit large internal fields and strong gradients that substantially amplify EDM signals compared to atoms~\cite{Safronova2018,AlarconElectric2022}. Recent progress in bringing molecular complexity under control~\cite{Cairncross2019, Dem24} has enabled $>\!100\times$ more precise bounds on the electron EDM in the last decade~\cite{Roussy2023,ACME2018}, significantly constraining a variety of new physics models at or beyond the direct detection reach of particle colliders~\cite{CesarottiInterpreting2019, PanicoEFT2019, BrodElectric2021,  AlarconElectric2022, BahlConstraining2022, KleyElectric2022, KumarElectric2024 , Arduequivalent2025}.}

Recent molecular EDM searches have focused primarily on leptonic EDMs{, due to both} their simpler hyperfine structure {and straightforward theoretical interpretation  compared to composite hadrons. However, with recent progress on the production and study of radioactive molecules~\cite{Garcia2020, Arr24, ConnProduction2025}, state-of-the-art measurements of hadronic CP-violating moments} are now within reach, { targeting orders of magnitude improvements in CPV sensitivity. Exploring the hadronic sector provides access to a broader range of new physics sources, beyond those constrained by lepton EDM measurements alone, including sensitivity to the strong CP problem~\cite{EngelElectric2013 ,Chupp2019}.}


{Specifically}, radioactive molecules containing pear-shaped nuclei \cite{Gaffney2013} offer a promising new platform for precision measurements {of fundamental symmetry violation~\cite{Arr24}.}
Certain heavy radioactive nuclei such as $^{223}\mathrm{Fr}$, $^{225}\mathrm{Ra}$, $^{227}\mathrm{Th}$, and $^{229}\mathrm{Pa}$ are either confirmed or suspected to exhibit octupole deformation and nearly-degenerate ($\Delta E\lesssim100~\text{keV}$), opposite-parity nuclear states, which can enhance CPV nuclear properties by more than three orders of magnitude when compared to stable nuclei \cite{Chupp2019,Auerbach1996}. {By embedding an octupole-deformed nucleus in a polar molecule, the additional molecular enhancement further amplifies observable CPV (see Figure~\ref{fig:amplifier}), yielding energy shifts that can be more than six orders of magnitude larger than in stable atoms \cite{Arr24, Safronova2018}.} Moreover, the rich internal structure of molecules can enable internal state-based co-magnetometry and quantum control protocols that reduce systematic uncertainties and increase effective sensitivity \cite{HutzlerPolyatomic2020, And23, Dem24, TakahashiEngineered2025}. Together, these features make radioactive molecules {uniquely attractive for researching fundamental symmetries} beyond the reach of current colliders~\cite{Pospelov05, EngelElectric2013, Safronova2018, Chupp2019, AlarconElectric2022, Arr24}.




\section{Symmetry Violating Nuclear Properties}\label{sec:probes}



{In atoms and molecules, direct nuclear EDMs,} {induced by $\bar \theta$ or new physics,} {are partially screened by the orbiting electrons, a result known as Schiff's theorem~\cite{Schiff1963}. Instead, hadronic CPV experiments probe the nuclear Schiff moment $\mathbf{S}$, a residual, CPV vector moment arising from relativistic dynamics and finite nuclear-size effects that evade electronic screening \cite{Flambaum2002, Liu2007, EngelNuclear2025}.}  CPV in atoms and molecules can also be observed via an electron EDM or nuclear magnetic quadrupole moment, though here we focus on the Schiff moment due to the particular nuclear enhancement discussed shortly. 

The nuclear Schiff moment is sensitive to the deviation of a nucleon's radial coordinate, $r^2$, from the mean-squared nuclear charge radius $R_{ch}^2$. This deviation from symmetry combines with the nucleon's charge $q$ or permanent EDM $\mathbf{d}$ to contribute to the Schiff moment, formally defined as~\cite{EngelNuclear2025} 
\begin{eqnarray}
    \mathbf{S} = \mathbf{S}_{ch} + \mathbf{S}_{n} + \ldots=\sum^A_{i=1} \left( \frac{1}{10} \left(r_i^2 - \frac{5}{3} R_{ch}^2 \right) q_i \mathbf{r}_i + \frac{1}{6} \left( r_i^2 - R_{ch}^2\right) \mathbf{d}_i \right) +\ldots, \label{eq:schiff_micro}
\end{eqnarray}
where $A=N+Z$ is the mass number of the nucleus, with $N$ and $Z$ the number of neutrons and protons respectively, $i$ labels each nucleon, the sum is taken over all nucleons, and $\ldots$ denotes higher-order contributions~\cite{Liu2007}. The charge term $\mathbf{S}_{ch}$ receives direct contributions from the protons only, while the nucleon term $\mathbf{S}_{n}$ contains contributions from permanent EDMs of individual nucleons. For large $A$ the charge term is expected to dominate, as there are typically only $\mathcal{O}(1)$ unpaired nucleons that can contribute to the sum over nucleon EDMs.

Since $\mathbf{S}$ is a vector, by the Wigner-Eckart theorem the associated observable must lie along the nuclear spin $\langle \mathbf{S}\rangle \equiv S_z = \mathbf{S}\cdot \hat{z}_N \propto \langle \mathbf{S} \cdot\mathbf{I}\rangle$, where $\hat{z}_{N}$ is along the quantization axis for the nuclear angular momentum $\mathbf{I}$. On the other hand, such a correlation of a P-odd vector $\mathbf{S}$ with a T-odd spin $\mathbf{I}$ can only exist if there is P and T violation, equivalent to CPV, in the nuclear Hamiltonian. Indeed, $S_z$ can be expanded in perturbation theory as:
\begin{equation}
    S_z = \sum_{e} \frac{\langle g|\mathbf{S}|e\rangle\langle e|H_{CPV}|g\rangle}{E_{g}-E_{e}} + c.c., \label{eq:schiff_pert}
\end{equation}
where the sum is taken over opposite parity excited states $|e\rangle$  with energy $E_e$. Here, $H_{CPV}$ is the CP-violating portion of the nuclear Hamiltonian~\cite{EngelElectric2013, deVriesParity2020}, discussed further in section~\ref{sec:reach}. The $H_{CPV}$ interaction slightly polarizes the ground state $|g\rangle$ by mixing the opposite-parity excited nuclear states, causing a CP-odd orientation of the intrinsic Schiff moment, collinear with $\mathbf{I}$, that can be observed in the lab-frame. Owing to both the near cancellation of terms in eq.~\ref{eq:schiff_micro}, and the sum over nuclear excited states in eq.~\ref{eq:schiff_pert}, nuclear theory calculations of Schiff moments require state-of-the-art methods~\cite{Dobaczewski2018, EngelNuclear2025, ZhouEffects2025}. 

The sensitivity of nuclear Schiff moments to new physics is dramatically enhanced in certain nuclei exhibiting static octupole shape deformation ($|\beta_3| > 0$) ~\cite{Auerbach1996, EngelNuclear2000, EngelElectric2013, Dobaczewski2005, Dobaczewski2018,EngelNuclear2025}. {The rotational states of such nuclei are described by symmetric top wavefunctions $|I,K\rangle$, with $K=\mathbf{I}\cdot \hat{z}_{N}$, which split into nearly-degenerate parity doublet states, $|\pm\rangle \propto |I,K\rangle\pm|I,-K\rangle$, similar to the description of nearly-degenerate, opposite-parity molecular states~\cite{ButlerIntrinsic1996,ButlerPearshaped2020}.} The Schiff moment enhancement arises from two key features in the perturbative expression of eq.~\ref{eq:schiff_pert}: the large intrinsic Schiff moment {$\langle I,K|\mathbf{S}|I,K\rangle$} in the body-fixed (intrinsic) frame due to collective effects, and the small energy splitting between opposite-parity states, { e.g., $\Delta E_\pm \approx 50$~keV in octupole-deformed $^{225}$Ra ~\cite{ButlerIntrinsic1996}. Parity doublets allow for the unique perturber approximation, reducing the sum over $|e\rangle$ in eq.~\ref{eq:schiff_pert} to a single opposite parity excited state, simplifying calculations. The expression for $S_z$ in octupole-deformed nuclei is then approximately given by\cite{EngelNuclear2000,Flambaum2002}}
\begin{eqnarray}
    && {S_z \approx 2\frac{\langle +|\mathbf{S}|-\rangle\langle -|H_{CPV}|+\rangle}{\Delta E_\pm}} \propto \frac{\beta_2 \beta_3^2 Z A^{2/3}}{\Delta E_{\pm}},
    \label{eq:schiff_scaling}
\end{eqnarray} 
{ where we have introduced the scaling of $S_z$ derived from a particle-rotor approximation of $H_{CPV}$ in static octupole-deformed nuclei.} Here, $\Delta E_{\pm}$ is the energy splitting between opposite-parity nuclear states, and $\beta_2$ and $\beta_3$ are the nuclear quadrupole and octupole deformation parameters~\cite{ButlerIntrinsic1996}. {The deformation parameters describe, in the intrinsic frame, the radius of the axially symmetric charge distribution as a function of the polar angle $\theta$, given in terms of spherical harmonic functions by $R(\theta) = R_{ch} \left(1+\sum_\ell \beta_\ell Y_{\ell,0}  (\theta,\phi) \right)$.}

{ From the scaling in eq.~\ref{eq:schiff_scaling}, the Schiff moment is enhanced compared to spherical nuclei, such as $^{199}$Hg, by a factor of $\mathcal{O}(1000)$, with specific values depending on nuclear structure details~\cite{EngelNuclear2000, EngelElectric2013, EngelNuclear2025}. In general, eq.~\ref{eq:schiff_scaling} is approximate; state-of-the-art calculations use EFTs to parameterize CPV interactions in the nucleus, and then solve the many-body quantum problem with density functional theory (DFT) techniques~\cite{Dobaczewski2005, Dobaczewski2018, ZhouEffects2025}, with some recent progress also ongoing in lighter nuclei using \textit{ab initio} nuclear theory~\cite{deVriesParity2020, GnechElectric2022, HergertGuided2020}. While we do not consider them further, we note that vibrationally octupole-deformed nuclei (with  $\langle \beta_3\rangle=0$ and $\langle \beta_3^2\rangle>0$) can potentially exhibit Schiff enhancements due to octupole collectivity~\cite{EngelNuclear2000, ButlerOctupole2016,FlambaumEnhanced2019}. However, due to much larger parity splitting in these systems~\cite{ButlerObservation2019, ButlerEvolution2020,ButlerPearshaped2020}, the unique perturber approximation breaks down, resulting in order-of-magnitude theoretical uncertainties~\cite{FlambaumEnhanced2019, SushkovSchiff2024} that would benefit from further investigation.} 

\begin{figure}[t]
\centering
\includegraphics[width=1\textwidth]{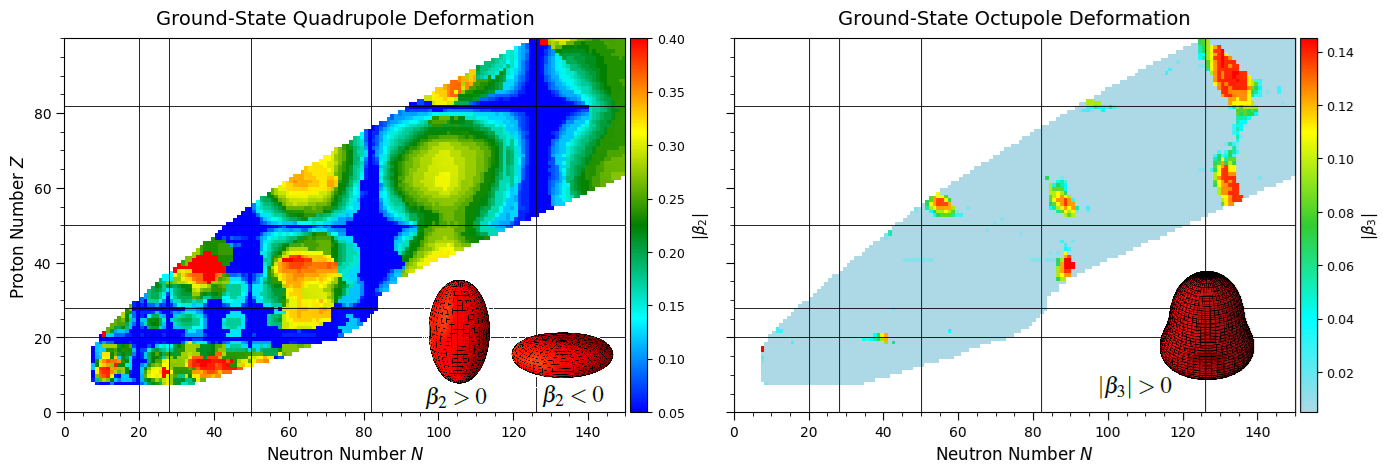}
\caption{Quadrupole and octupole deformation across the nuclear chart. The deformation parameters $\beta_2$ and $\beta_3$ characterize deviations of a nucleus from spherical symmetry. The quadrupole parameter $\beta_2$ corresponds to elongated (prolate) or flattened (oblate) shapes, respectively. The octupole parameter $\beta_3$ describes pear-like shapes that do not exhibit reflection symmetry. Figure modified from \cite{FRDM12}.\label{NuclearChart} }
\end{figure}

Figure~\ref{NuclearChart} provides an overview of nuclei exhibiting static quadrupole ($|\beta_2|>0$) and octupole deformations ($|\beta_3|>0$). While quadrupole deformation is common in nuclei, octupole deformation occurs in specific places of the nuclear chart. In the nuclear shell model, octupole deformation originates from {$r^3 Y_{3,0}$} interactions between nuclear levels separated by $3\hslash$ of angular momentum near the Fermi surface~\cite{EngelNuclear2000, ButlerOctupole2016,ChenMicroscopic2021}. {This occurs} primarily in nuclei whose proton or neutron numbers ($Z$ or $N$) are close to 34, 56, 88, or 134, just beyond spherical magic numbers~\cite{CaoLandscape2020}. The deformation is especially prominent in heavy, exotic nuclei near $Z \approx 88$ and $N \approx 134$ (e.g., isotopes of Fr, Ra, Ac, Th, and Pa)~\cite{ButlerPearshaped2020}. However, the significant nucleon imbalance in these isotopes renders them radioactive, typically limiting their lifetimes to a few days or less. {Nonetheless, significant progress has been made recently on the production and spectroscopy of molecules containing octupole-deformed nuclei~\cite{Garcia2020, Arr24, ConnProduction2025}, providing a unique opportunity for next-generation CPV searches. }

\section{Effective Field Theories and Energy Reach}\label{sec:reach}

\begin{figure}[!tp]
\centering
\includegraphics[width=\columnwidth]{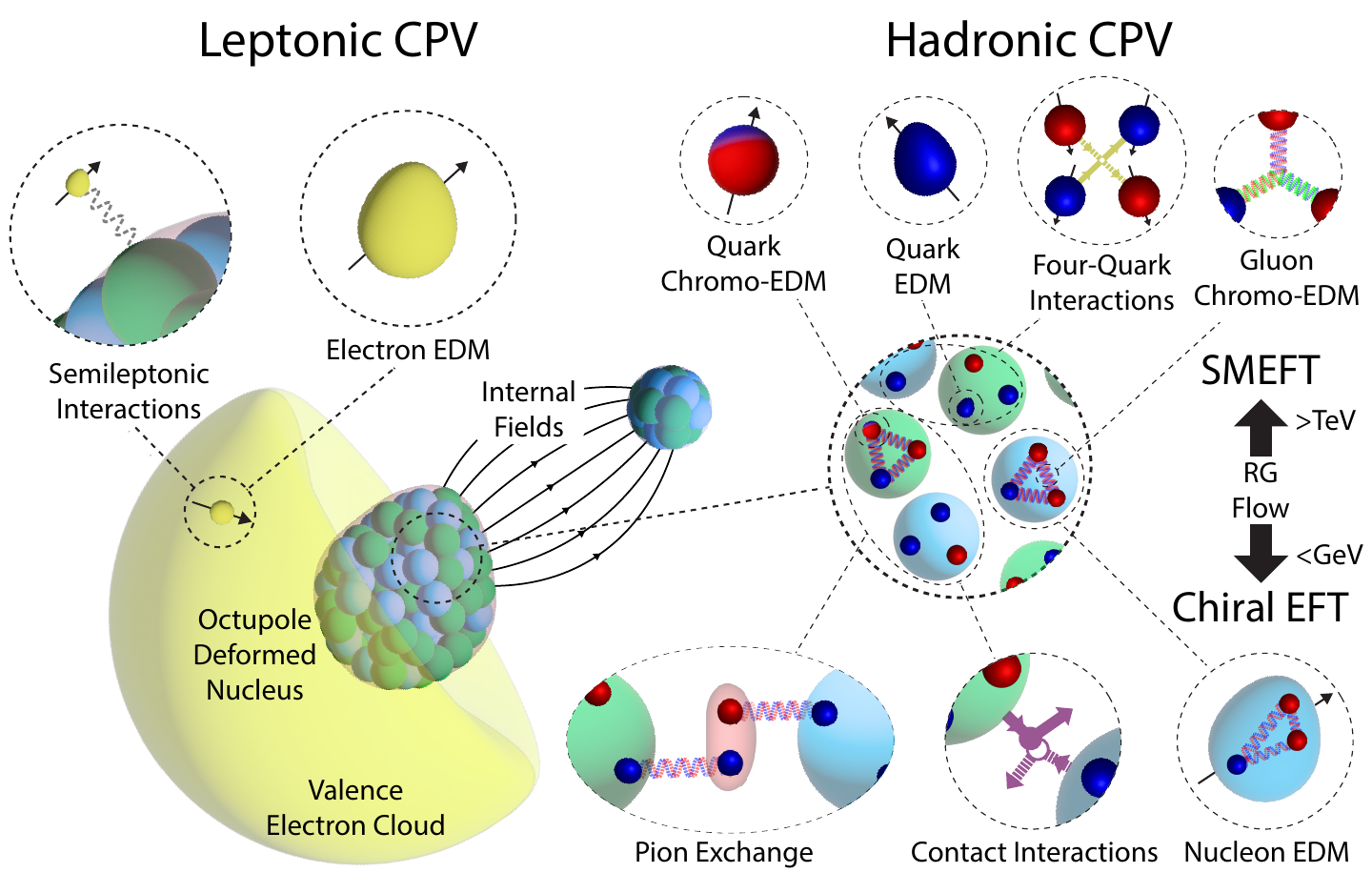}
\caption{
Schematic representation of a diatomic molecule and {sources of CP-violation (CPV) among its constituents}. Electrons interact {with the nucleus} primarily via electromagnetic and weak forces, {and electron EDM experiments primarily probe leptonic CPV.} Inside nuclei, additional complex interactions occur among protons and neutrons, as well as their fundamental building blocks---quarks and gluons---through a combination of the electromagnetic, weak, and strong forces. { At high energies, these microscopic QCD sources of CPV are captured by SMEFT, and connected via Renormalization Group (RG) flow to low-energy $<$GeV chiral EFT interactions. See text for detailed descriptions of the interactions and EFTs. Octupole-deformed nuclei inside molecules with large internal fields provide enhanced sensitivity to a multitude of sources of hadronic CPV, most prominently CPV pion-exchange~\cite{Cairncross2019,Arr24}.}
\label{fig:molecule}}
\end{figure}

To interpret searches for CPV moments as constraints on new physics, we must connect low-energy
($<\text{GeV}$, ``IR'') observables to their microscopic, high-energy
($>\text{TeV}$, ``UV'') sources. This connection is made through a tower of effective field theories
(EFTs) that systematically bridge vastly different energy scales~\cite{GeorgiEffective1993,
ManoharIntroduction2020,HammerNuclear2020,Isidoristandard2024}. EFT approaches enable systematic,
global analyses of CPV using EDM bounds from nucleons, atoms, and molecules~\cite{DegenkolbGlobal2025,
GaulGlobal2024} as well as combined analyses with collider experiments~\cite{Ram21,BahlConstraining2022,
BrodGlobal2022}. Two key EFT frameworks are the Standard Model EFT (SMEFT)~\cite{Isidoristandard2024,
AebischerSMEFT2025} in high-energy physics and chiral EFT ($\chi$EFT)~\cite{HammerNuclear2020,
deVriesParity2020,EngelNuclear2025} in nuclear physics. SMEFT describes how generic new particles and
forces at $>\text{TeV}$ energies modify interactions of Standard Model fields at and below the
electroweak scale ($\lesssim\text{TeV}$), while $\chi$EFT describes low-energy, non-perturbative QCD
at $<\text{GeV}$ energies in terms of composite nucleons and pions.  A schematic overview of EFT sources of molecular CPV is shown in Fig.~\ref{fig:molecule}.

While EFTs provide a systematic framework for connecting CP-violating observables to microscopic
sources of new physics, they represent only one layer in the full theoretical interpretation chain
relevant for radioactive molecule experiments. On the low-energy side, $\chi$EFT interactions must be
embedded in realistic nuclear many-body calculations to determine nuclear moments such as Schiff
moments, particularly in heavy, octupole-deformed nuclei. In parallel, relativistic atomic and molecular
structure calculations are required to translate these nuclear moments into experimentally observable
energy shifts, through the determination of electronic enhancement factors and internal effective fields.
In practice, estimates of experimental sensitivity for a given radioactive molecule rely on the combined
input of EFT matching, nuclear structure theory, and molecular electronic structure calculations, often
well in advance of experimental measurements. Continued progress across all three areas is therefore
essential for both the interpretation and design of next-generation searches for CP violation.
For heavy nuclei, state-of-the-art calculations
employ nuclear density functional theory~\cite{Dobaczewski2018,ZhouEffects2025}, while \textit{ab initio} approaches remain limited to lighter
systems~\cite{DekensUnraveling2014,Froeseinitio2021,HergertGuided2020}. At the molecular level, relativistic many-electron calculations determine the electronic
enhancement factors and internal effective fields that convert a given Schiff moment into a measurable
frequency shift~\cite{ChenRelativistic2024}.
After parameterizing nuclear Schiff moments in terms of $\chi$EFT and SMEFT
coefficients, SMEFT can be used to estimate the energy reach of hadronic EDM experiments.
Owing to the large number of CPV operators that couple to quarks and gluons, hadronic EDM searches
probe a broader region of SMEFT parameter space than purely leptonic measurements. 

\begin{figure}[!tp]
\centering
\includegraphics[width=1\textwidth]{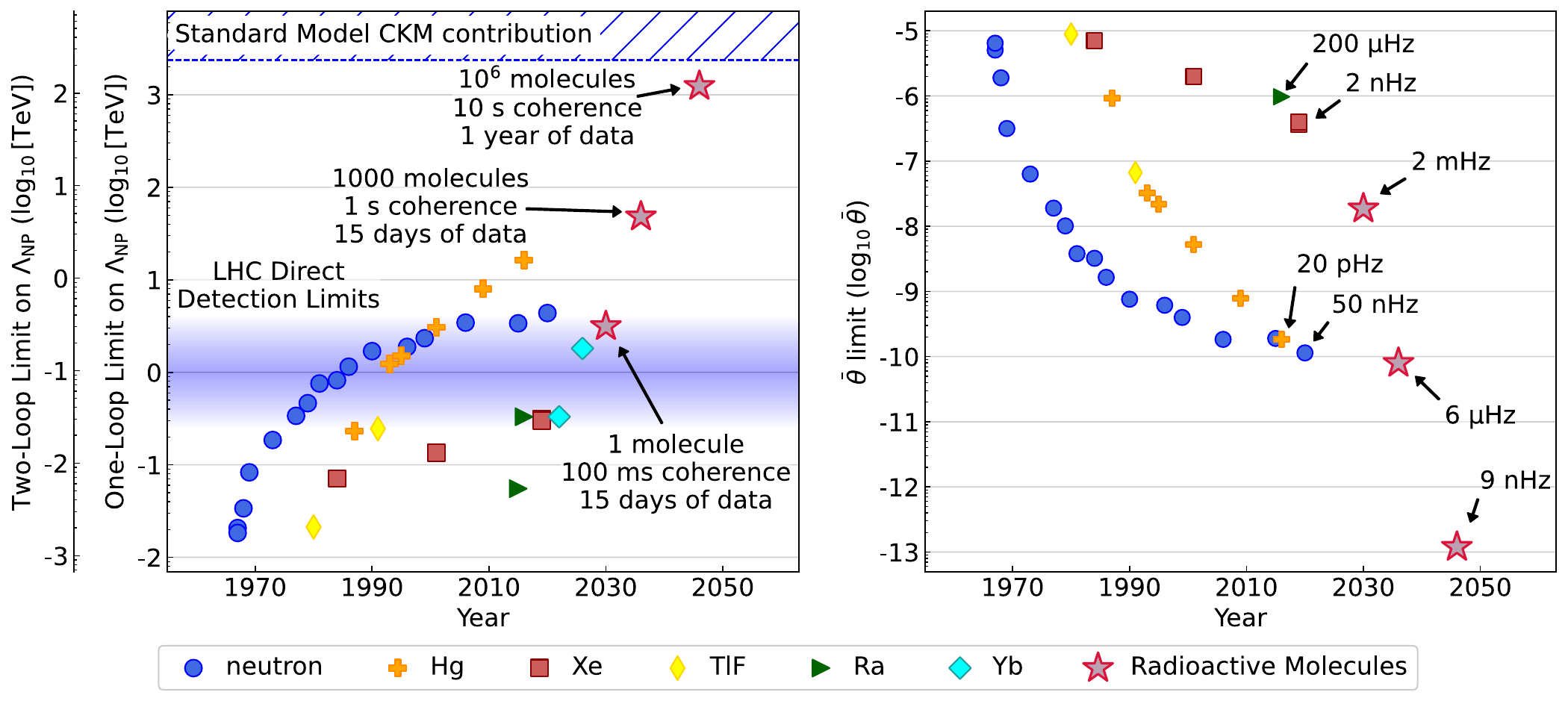}
\caption{\textbf{Hadronic EDM Limits.} Prospects for probing new physics with radioactive molecules {(pink stars)}, in comparison to current LHC limits {(horizontal faded band) and previous hadronic EDM measurements. Hadronic enhancement factors in eq.~\ref{eq: schiff_enhance_sources} (see also Supplemental Material) are estimated by geometric means of the values in Ref.~\cite{EngelNuclear2025}. Left: one- and two-loop limits on the new physics energy scale, $\Lambda_\text{NP}$, obtained from a single-source approximation of $\bar g_1$ from an isovector quark CEDM $\tilde d_{1}$. See Ref.~\cite{AlarconElectric2022} for more details. LHC constraints taken from Refs.~\cite{Kwon2025,ATLAS2025,Sekmen2025}. The equivalent signal from the Standard Model CKM contribution is shown with the hatched region. Right: limits on $\bar \theta$ from a single-source approximation considering only CPV pion-exchange $\bar g_0$. The best limits from $^{225}$Ra, $^{129}$Xe, $^{199}$Hg, and the neutron, as well as radioactive molecule projections, are labeled with their frequency precision, contextualizing differences in sensitivity. For $\bar \theta$, the expected Standard Model CKM contribution~\cite{Chupp2019, DragosConfirming2021} at $\bar \theta \sim 10^{-16}$ and the QCD axion oscillation scale~\cite{deVriesUncovering2021} at $\bar \theta_{rms} \sim4\times 10^{-19}$ both lie outside of the plotted range. All calculations assume 1 second experimental cycle dead-time. Compiled measurements and constraints are available at Ref~\cite{JayichGitHub2025}.}
}
\label{fig:prospects}
\end{figure}

{Using dimensional analysis arguments, we can estimate the energy reach of hadronic EDM measurements with radioactive molecules containing octupole-deformed nuclei. For simplicity, we focus on the fermionic EDMs and chromo-EDMs, which can be estimated as~\cite{CesarottiInterpreting2019, Chupp2019,AlarconElectric2022}:}
\begin{equation}
d_f \sim q_\text{f} \sin{\phi_\text{CP}}\left(\frac{g^2}{16\pi^2}\right)^\ell \frac{m_\text{f}}{\Lambda_\text{NP}^2}, \label{eq:edm_est}
\end{equation}
where $q_\text{f}$ and $m_\text{f}$ are the fermion's (color) charge and mass, $\phi_\text{CP}$ is the CP-violating phase {characterizing the complex Wilson coefficient\footnote{The contribution to the CP-even magnetic dipole is scaled by $\cos{\phi_\text{CP}}$.},} $\ell$ is the number of loops in the Feynman diagram that gives rise to the EDM, and $g$ is the strength of the coupling constants in the loop. { Loops involving SM interactions receive factors of $g_\text{EM}\sim 0.3,~g_\text{W}\sim 0.6,~g_\text{S}\sim1$, and typically $\sin(\phi_\text{CP})$ and $g_\text{BSM}$ are taken to be $\mathcal{O}(1)$ quantities without additional fine-tuning. Tree-level ($\ell=0$) diagrams have the highest energy reach but are sensitive to only a few BSM interactions, while one- and two-loop ($\ell=1$ and $\ell=2$) diagrams incorporate wider operator-reach at the cost of $\mathcal{O}(10^2)$ to $\mathcal{O}(10^4)$ loop suppression. While two-loop bounds generically have suppressed energy reach compared to one-loop bounds, two-loop processes can be enhanced by heavy quark contributions and renormalization mixing, relevant for constraining various new physics scenarios~\cite{Pospelov05,CesarottiInterpreting2019,PanicoEFT2019}.}

Figure \ref{fig:prospects} illustrates the one-loop energy reach {probing $\Lambda_\text{NP}$} and the value of $\bar\theta$ constrained by prior hadronic EDM measurements using systems such as the neutron, {atomic} Hg, Xe, {and Ra,} and {molecular} TlF. {The leading bounds on hadronic CPV arise from EDM} measurements of the neutron~\cite{Abel2020} and Hg atom~\cite{GranerReduced2016}, {already probing energy scales on the order of $\sim$10~TeV in the single-source approximation}.{ To compare with the potential reach of radioactive molecules, we calculate the EDM limit assuming measurements at the Standard Quantum Limit (SQL)~\cite{AlarconElectric2022}:
\begin{equation}
    \delta d_f \simeq 
    \underbrace{\frac{\hslash}{\tau \sqrt{N R T}}}_\text{SQL}
    \times
    \underbrace{\frac{1}{\langle I_z\rangle W_S}}_\text{Molecular}
    \times
    \underbrace{\frac{1}{S_z \left(\bar g_0  (\bar \theta \ldots),\bar g_1(\tilde d_1 \ldots), \ldots\right)}}_\text{Nuclear} \label{eq: schiff_enhance_sources}
\end{equation}
where the first factor describes the SQL obtained from Ramsey measurements, given a coherence time $\tau$ and $N R T$ detected molecules, with $N$ molecules per experiment cycle, $R$ the cycle repetition rate, and $T$ the total experiment run-time\footnote{Note when the repetition rate is limited by the coherence time, $R \propto \tau^{-1}$, and the coherence time scaling is reduced to $\delta d_f \propto \tau^{-1/2}$.}. The molecular enhancement term is given by the product of the polarization of nuclear spin along the internal molecular axis, $\langle I_z\rangle$, and the calculated electronic enhancement factor $W_S$ describing the internal molecular fields near the nucleus~\cite{ChenRelativistic2024}. Finally, the nuclear enhancement term (see Section~\ref{sec:probes}) describes the scaling of the Schiff moment in terms of the underlying LECs, which can then be related to underlying Wilson coefficients.}

{The figure shows the potential constraints that could be set using molecules with octupole-deformed nuclei, labeled by number of molecules per experiment cycle, coherence time, and total measurement duration. For reference, state-of-the-art trapped molecule experiments have achieved $>10^4$ laser-cooled CaF molecules in a conservative optical trap at 14 \textmu K~\cite{YuConveyer2024}, and seconds-scale coherence in optically trapped, ultracold RbCs molecules. Importantly, there is a pathway to further improve bounds by leveraging the toolbox of molecular quantum control~\cite{Cairncross2019, Dem24}. Advanced measurements with radioactive molecules} could potentially achieve constraints beyond the 1000 TeV scale, thereby covering the discovery window between the limits set by direct searches at the LHC and the known sources of CPV within the Standard Model. The parameters required to reach the Standard Model contribution have already been demonstrated in state-of-the-art experiments with cold, stable atoms and molecules{~\cite{Safronova2018, Dem24}}; applying these techniques to radioactive molecules promises to have a significant impact on searches for new symmetry-violating physics.

\section{Experimental {Techniques} and Challenges}
\label{sec:tools}
\begin{figure}[!tp]
\centering
\includegraphics[width=1\textwidth]{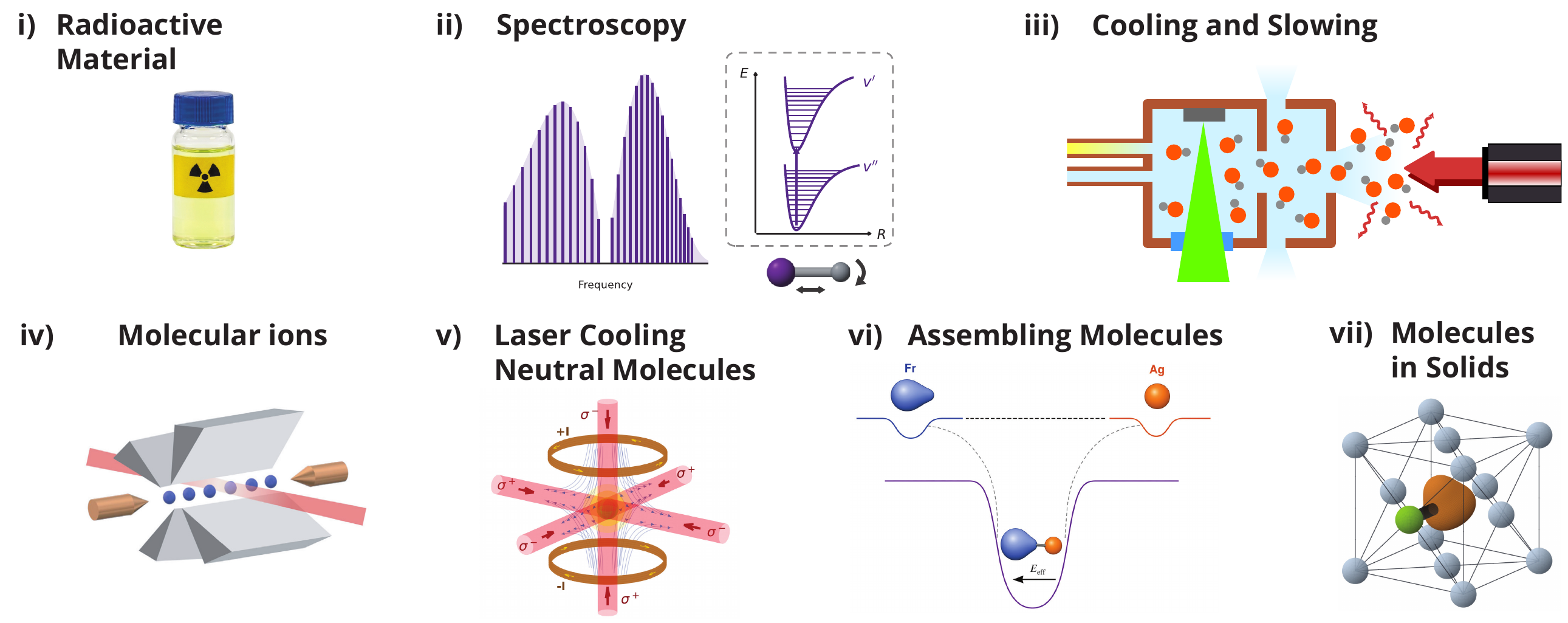}
\caption{\textbf{Emerging tools for precision measurements with radioactive molecules.} Clockwise from top left: i) Cartoon of a vial containing a compound of radioactive material obtained from an isotope production facility~\cite{KhasanovMajor2025,Abel19,CollinsUS2020}; ii) Initial spectroscopy to determine molecular structure is essential for any high-precision experiment~\cite{Garcia2020, Yan23,Udrescu2024,wilkins24,Athanasakis2025}; iii) Cryogenic buffer gas cells~\cite{Hutzler2012} for radioactive molecule production~\cite{ConnProduction2025}; iv) Radiofrequency trap used for the study of molecular ions~\cite{Roussy2023, Fan2021RaPoly,Fan23,Rosch2016} (figure from Ref.~\cite{Yan23}); v) Laser cooling and trapping of neutral molecules in a magneto-optical trap~\cite{Fitch2021Review,Kozyryev2017,YuConveyer2024, Lasner2025SrOHMOT};  vi) Ultracold molecules produced by assembling ultracold atoms~\cite{KlosProspects2022,MarcCandidate2023}; vii) Solid-state matrix isolation setup for study of molecules in solids~\cite{edm3,pa229,ian23,rv23}. \label{Experiments}}
\end{figure}

Radioactive molecules containing {octupole-deformed} nuclei are {often} produced in extremely {trace} quantities and {are typically reactive or refractory.} As a result, major challenges are present in precision measurement with these molecules. Experiments with any significant amount of radioactive material prove difficult, requiring special consideration in storage, handling, {disposal, and experimental} design. This scarcity also {requires} highly efficient {chemistry}, trapping, spectroscopy, and measurement techniques, {at a level} not previously necessary for stable species. 
Figure \ref{Experiments} presents the experimental developments underway for creation, preparation, and precision measurement with radioactive molecules.

\noindent \textbf{i) Production of radioactive isotopes.}
The growing interest in studying unstable nuclei for both {foundational} physics and practical applications has driven the development of numerous major radioactive beam facilities worldwide~\cite{KhasanovMajor2025}, such as FRIB {at MSU} (United States), ISOLDE at CERN (Europe), {RIBF at RIKEN (Japan)}, and {ISAC at} TRIUMF (Canada). Experimental programs focused on radioactive molecules are being developed at these facilities and are expected to be operational in the coming decade~\cite{Arr24}. These facilities employ various nuclear reactions to produce and efficiently extract isotopes of interest in both atomic and molecular forms. ISOLDE and TRIUMF provide chemically separated, high-purity atomic and molecular beams, while FRIB generates a broader range of exotic nuclei through in-flight {fragmentation~\cite{TarasovObservation2024}.} The latter is more suitable for refractory elements such as Th and Pa, both of special interest for symmetry violation searches. A new isotope harvesting initiative at {FRIB~\cite{AbelIsotope2019}} aims to provide access to short- and long-lived radioisotopes, such as $^{225}$Ra, $^{225}$Ac, and $^{229}$Th, which could serve as valuable sources of rare isotopes {without requiring dedicated beam-time. Harvested isotopes of interest} can be purified and transported for precision experiments and {rapid prototyping} at university laboratories.

In addition to beam facilities, major nuclear physics laboratories, such as {ORNL~\cite{PattonMinor2019,CollinsUS2020}} (United States), specialize in the production and separation of actinide and transuranium nuclei, to facilitate construction of offline sources. For instance, experiments laser cooling and {trapping~\cite{SprouseLaser1997} atomic} $^{221}$Fr and $^{225}$Ra utilized sources with the parent isotope $^{225}$Ac and $^{229}$Th, {respectively~\cite{TandeckiJINST2014, Fan23}}. In these sources, the daughter products are collected either via evaporation of the daughter through heating, or via alpha-decay recoil which ejects the daughters from the surface of the sample. While offline sources {typically provide less instantaneous flux than} dedicated facilities, {in many cases they offer a complementary alternative compared to the costs and scheduling limitations of facility beam-time. Additionally, for species with long-lived parent isotopes, offline sources can operate for years to decades without need for renewal. Overall, offline sources can provide a flexible, potentially long-term solution enabling various experiments at university laboratories. Mature experiments can then be adapted to perform online measurement runs at beam facilities, taking advantage of high-flux radioisotope production and advanced radioactive material handling capabilities. }

\noindent \textbf{ii) Spectroscopy.}
Precision spectroscopy of short-lived radioactive molecules has become possible in the last few years thanks to the development of highly sensitive production and detection schemes \cite{Garcia2020}. Starting with the production of radioactive molecular ions RaF$^+$, followed by their subsequent neutralization into RaF, overall efficiency of more than 10$\%$ has been demonstrated, with molecular fluxes on the order of 10$^7$ molecules per second \cite{Garcia2020, Udrescu2021}, enabling studies of the short-range electron-nucleus interactions \cite{wilkins24}, and providing an experimental determination of a laser cooling scheme for RaF \cite{Udrescu2024,Ath23,Ath24}. This progress has motivated the emergence of a new research area at various laboratories across Europe and North America \cite{Arr24}, {with experiments already extended to other actinide molecules such as AcF \cite{Athanasakis2025}}. However, these experiments were performed using a hot, energetic beam of molecules, with kinetic energies of about 40 keV and internal temperatures exceeding 1000 K. 

A primary challenge lies in the efficient production of slow, cold radioactive molecules with sub-Kelvin temperatures, which are essential for further precision experiments and enabling efficient laser-cooling techniques. Recent progress includes production and spectroscopy of cold $^{226}$RaOH, $^{226}$RaOD, and $^{226}$RaF, via cryogenic buffer gas cooling~\cite{Hutzler2012} and laser-driven chemistry~\cite{JadbabaieEnhanced2020}, with yields of up to $\sim10^{10}$ molecules per pulse at temperatures between 4--7~K \cite{ConnProduction2025}. These techniques can be readily applied to obtain laser spectra of a wide variety of diatomic and polyatomic radioactive species in the near future. 




\noindent \textbf{iii) Deceleration and Cooling.} Measurements with cold beams of molecules {are} a demonstrated method for performing spectroscopy with kHz or better resolution, as well as precision measurements of fundamental symmetries with $<$mHz precision.  The ACME experiment, for example, has used a cold beam of {neutral} ThO molecules to set limits on the electron's electric dipole moment \cite{ACME2018}.  In order to realize maximum sensitivity, these beam sources must be cold, slow, and bright.  A temperature of $\lesssim$10~K ensures that a relatively small number of internal states are populated, { concentrating population in the lowest rotational and hyperfine quantum states.} High brightness (flux per unit solid angle per second) ensures high counting statistics. Low forward velocity { not only} increases interaction time while maintaining a short beam line, {but also enables slowing and cooling into a trap.  For neutral molecules,} the leading approach is the cryogenic buffer gas beam\cite{Hutzler2012}, which can be used to create cold, slow, and bright beams of essentially any species, including those which are reactive or refractory. 
Molecular beams can be produced directly in a cryogenic buffer gas cell {by laser ablation of a precursor in the presence of buffer gas, which can also be combined with reagent gas flowed in via a heated capillary~\cite{Hutzler2012}. Alternatively,} cryogenic buffer gas cells can be used to slow initially hot and fast beams. This approach has been used {to produce cryogenic beams of atoms loaded from thermal ovens~\cite{EgorovBuffer2002}, and also used at radioactive ion beam facilities to stop GeV beams~\cite{KalejaPerformance2020,LundOnline2020}.} Neutral radium-containing species, which have been recently cooled in a cryogenic buffer gas cell~\cite{ConnProduction2025}, are currently being pursued for laser cooling and trapping~\cite{Fitch2021Review,Augenbraun2023PolyLCReview}.

\noindent \textbf{iv) Ion Trapping.}
Ion traps are well-suited to working with radioactive atomic and molecular species \cite{Dilling2018}. {Strong electromagnetic control relaxes requirements on low beam velocity, enabling rapid prototyping and production of free radicals using supersonic expansion cooling~\cite{Hutzler2012}. Further,} ion traps are deep, which makes it routine to hold particles continuously for months (or until they decay) \cite{Olmschenk2009a}. The trapping mechanism only depends on the charge and mass of the molecule, making ion traps agnostic { at first order} to complex internal molecular {structure~\cite{Rosch2016}. Additionally, higher-order sensitivities can provide an avenue for readout via Stern-Gerlach-like effects~\cite{Hermanspahn1999}.} In addition to the long hold times, ion traps may be used to efficiently work with a few particles, often realizing precision measurements with only a single trapped ion \cite{Rosenband2008}.

Because ion traps are insensitive to internal structure, it is straightforward to simultaneously hold different elements in the same trapping potential, {as long as the charge-to-mass ratios are not too different.}  Co-trapping laser-coolable atoms with molecular ions provides opportunities for controlling the molecules with the atoms via the shared motional modes of the ion trap.  An early realization~\cite{Schmidt2005} that leveraged these strengths was quantum logic spectroscopy (QLS) of Al$^+$ with co-trapped Be$^+$.  In this experiment Be$^+$ cooled a motional mode of the ion pair to the ground state, and then was used to perform high fidelity readout of Al$^+$ after it was probed with a spectroscopy pulse.  QLS was developed because Al$^+$, like molecules, does not have good cycling transitions, and therefore this approach could also be used for molecular spectroscopy \cite{Schmidt2005}.   This led to QLS state detection of a molecule \cite{Wolf2016}, followed by state detection and state preparation of a molecule \cite{Chou2017}, and subsequent advances in controlling molecule rotational states \cite{Chou2020}. QLS could provide a means to very efficiently use a single (or few) particles that may be challenging to produce in large quantities, {especially helpful for trace radioactive species.}  Other efficient spectroscopy techniques have recently been developed that could be applied to polyatomic radioactive molecules \cite{Calvin2023}.

There have been recent advances in Paul ion trap technology, driven by both quantum information science \cite{Moses2023, xu20233} and precision measurement \cite{Chen2017}, including world-record eEDM measurements using HfF$^+$\cite{Cairncross2019, Roussy2023}. Precision science using heavy molecular ions with small rotational constants will likely require cryogenic systems to reduce population scrambling by blackbody radiation.  Cryogenic ion traps are seeing increased use due to advantages such as their very low pressures \cite{Pagano2019}.  For EDM searches molecules must be polarized with an electric field, but the molecular ions should not be pushed by the field out of the region where they are polarized.  This challenge has been addressed with rotating electric fields \cite{Roussy2023}, but there are opportunities to use other dynamic schemes \cite{Verma2020, Zhang2023a}. Additionally, experiments trapping radioactive molecular ions can leverage emerging techniques, such as QLS. The demonstration of quantum logic spectroscopy and coherent state control with CaH$^+$ molecular ions\cite{Chou2020} offers a compelling route for precision measurements with molecular ions like RaOH$^+$ or RaOCH$_3^{+}$\cite{Kozyryev2017,Yu2021RaOCH3,Fan2021RaPoly} or polycations like PaF$^{3+}$\cite{Zulch2022Cool}, where the long-lived electronic states provide natural shielding from blackbody radiation.

Penning ion traps are also intriguing for their use with radioactive molecules.  They have been used extensively with radioactive species \cite{Dilling2018}, as well as for precision measurements \cite{Fan2023c}.  In part due to their large magnetic field, Penning traps have been recently considered for measuring parity violation with radioactive molecules \cite{Karthein2024}.

\noindent \textbf{v) Laser Cooling and Trapping.} Performing laser cooling and trapping of molecules sensitive to fundamental {symmetries}, including radioactive molecules, is a major area of current work \cite{Arr24}.  Ultracold and trapped neutral species offer a pathway to {large numbers}, long coherence times, and extreme levels of control, and these have been important drivers in the recent major advances in ultracold atom and molecule science.  However, direct laser cooling of molecules is an active area of research \cite{Dem24}, and extending these techniques to radioactive molecules remains one of the major current challenges in the field.

In recent years, there have been major advances in direct laser cooling of molecules since the first experimental demonstration was shown in 2010 \cite{Shuman2010}, including optical dipole trapping, magnetic trapping, tweezer trapping, controlled interactions, quantum gates, and quantum control for fundamental symmetry-violation searches \cite{Fitch2021Review,Augenbraun2023PolyLCReview,Dem24}.  Molecules that can be laser cooled using known methods are fairly uncommon, but radium-containing species such as RaF~\cite{Isaev2010RaF,Garcia2020,Udrescu2024} and RaOH~\cite{Isaev2017,Zhang2023RaOH} are predicted to be laser-coolable, thereby offering a potential route toward integrating state-of-the-art quantum methods with radioactive nuclei bound in molecules.

Current efforts with stable molecules have pushed the frontier of laser cooling and quantum control for complex molecules. The laser cooling of triatomic species like CaOH~\cite{And23} and SrOH~\cite{Lasner2025SrOHMOT}, which offer better polarizability and systematic corrections compared to diatomics\cite{Kozyryev2017}, has opened exciting prospects for symmetry tests with radioactive analogs like RaOH\cite{Kozyryev2017,Isaev2017}.

\noindent \textbf{vi) Assembling molecules.} In addition to direct laser cooling, radioactive molecules could potentially be assembled from laser-cooled atoms. This has been a successful approach in ultracold {bi-alkali molecules~\cite{BohnCold2017}}, though these molecules tend to have low sensitivity to {fundamental symmetry-violation~\cite{Meyer2009}}, even if one of the species is heavy like Fr or Ra.  However, bonding {heavy alkali or alkaline-earth atoms with} Ag results in a large molecular enhancement of CPV sensitivity~\cite{Fle21,KlosProspects2022,MarcCandidate2023,MarcSemi2025}, and experiments are currently under {development~\cite{Vayninger2025} to produce ultracold $^{223}$FrAg.}
Likewise, optical tweezer arrays have emerged as a powerful tool for assembling {individual} molecules from ultracold {atoms~\cite{LiuBuilding2018,CairncrossAssembly2021}} and may provide a route to create ultracold, radioactive {molecules from trace samples of radioactive atoms. }

\noindent \textbf{vii) Solid State Systems.} {Additional promising directions involve the use of noble gas solids to trap and orient molecules \cite{edm3} and atomic ions embedded in optical crystals \cite{pa229,rv23} or diamonds \cite{ian23} where they are exposed to large local electric fields. These solid state systems can accommodate number densities as high as $10^{15}/\mathrm{cm}^3$, far exceeding those of conventional laser and ion traps, while remaining sufficiently dilute to allow nuclear spin coherence times on the order of a second or longer \cite{vv48,ka53}. In a suitably chosen host solid, as first pointed out by Royce and Bloembergen \cite{rb63} in the context of paramagnetic ions in corundum, the guest species occupies pairs of trapping sites where the effective internal electric field felt by the nucleus is equal in magnitude but opposite in orientation \cite{mims} making the rejection of common-mode systematic effects feasible. Recent progress using matrix isolation techniques, where molecules are trapped in inert gas solids, offers a unique platform for precision spectroscopy and quantum control \cite{edm3}. The ability to efficiently use small samples, the prospect of oriented molecules in the matrix, and the long coherence times could be particularly advantageous for experiments with rare radioactive species. }


\section{Summary and Outlook}
\label{sec:summary}

Radioactive molecules offer a promising platform {to probe new} physics beyond the Standard Model. The field is rapidly progressing due to advances in experimental methods for producing and manipulating molecules {with radioactive nuclei, alongside theoretical developments in nuclear physics and QCD that improve calculations of nuclear properties and sensitivities.} In the coming years, experiments combining exotic nuclei with quantum-controlled molecules promise to improve current constraints on {CPV hadronic physics} by more than two orders of magnitude.

{In addition to searches for CPV, precision studies of radioactive molecules are of interest across multiple subfields in physics and chemistry~\cite{Arr24}. Measurements of isotope shifts and hyperfine structure can probe nuclear size variations and electromagnetic moments in nuclear matter far from equilibrium. While the vast majority of such experiments have been conducted using radioactive atoms~\cite{Yan23}, extending such studies to radioactive molecules is of interest, motivated by the enhanced sensitivity molecules can offer to study nuclear electroweak properties~\cite{Karthein2024} and their importance for astrophysics~\cite{Arr24}. Radioactive molecule spectroscopy is particularly useful to study elements that preferentially form molecules, or where atomic states have angular momentum structure that is insensitive to nuclear moments. Moreover, laboratory spectra of complex radioactive molecules are essential for unambiguously identifying such species in astrophysical environments, where they can be naturally produced~\cite{KaminskiAstronomical2018}. }

{There are also} exciting opportunities to further develop radioactive molecular techniques and expand the range of systems being studied. Improvements in molecular beam intensity, laser cooling, and ion trapping will enable experiments with even rarer molecular species. {Further, the integration of new approaches, from isotope harvesting to quantum-enhanced sensing to matrix isolation, is key to unlocking the full potential of radioactive molecules. As these techniques mature and are combined in new ways, they will enable a broad range of precision measurements with octupole-deformed nuclei. With unprecedented sensitivity to fundamental symmetry violation,} radioactive molecules can fully cover the discovery window (see Figure \ref{fig:prospects}), with the potential to unveil new sources of CP-violation, shed light on the strong CP problem, and probe physics {at the PeV scale.}

Realizing this scientific potential will require a concerted effort across traditional disciplinary boundaries. {Success will be enabled by experimental} advances in rare isotope production at accelerator facilities, laser cooling and molecular beam techniques, ion trapping and quantum control, {and ultracold assembly of molecules, with these efforts matched by progress in} relativistic many-body theory, nuclear structure models, and particle physics calculations. The multifaceted nature of {these challenges} presents opportunities for cross-community collaboration and interdisciplinary innovation.

As new experiments with radioactive molecules come online in the coming decade, transformational discovery may be at hand. Setting stringent limits on hadronic CP violation and uncovering the nature of {symmetry-}violating nuclear properties could reshape our understanding of the universe, potentially revealing new physics beyond the Standard Model. {
With many orders of magnitude separating current bounds and Standard Model predictions, any nonzero EDM measurement in the near future would be evidence of new physics, while continued null results further tighten bounds on hypothetical new particles and forces.} {In either outcome, the enhanced sensitivity and broad reach of precision measurements with radioactive molecules position them to play a leading role in exploring the fundamental laws of nature.}

\section*{Acknowledgments}
A.J. acknowledges support from the National Science Foundation (PHY-2402254). R.F.G.R acknowledges support from the U.S. Department of Energy, Office of Science, Office of Nuclear Physics under grants DE-SC0021176 and DE-SC0021179.  N.R.H. acknowledges support from the National Science Foundation (PHY-2309361 and CAREER PHY-1847550) and the Heising-Simons Foundation (2022-3361).  A.M.J. acknowledges support from the Department of Energy DE-SC0022034, the National Science Foundation PHY-2146555, and the Heising-Simons Foundation 2022-4066.  J.T.S acknowledges support from the U.S. Department of Energy, Office of Science, Office of Nuclear Physics, under contracts DE-SC0019015, DE-SC0019455, and DE-SC0025679. We thank David DeMille for useful discussions and his input for Figure 4 in our manuscript.

\bibliography{references}

\section*{Supplemental Material}

Here we provide some additional details about effective field theories (EFTs) and the interpretation of molecular CPV sources.
 Matching
between high- and low-energy regimes is achieved by integrating out heavy fields while retaining light
degrees of freedom~\cite{Isidoristandard2024}. The full EFT Lagrangian is given by an infinite sum over
operators of energy dimension $d$:
\begin{eqnarray}
\mathcal{L}_{UV}(\mu)\simeq\mathcal{L}_{EFT}(\mu)=
\mathcal{L}_{d\leq 4}(\mu) +
\sum^\infty_{d=5} \frac{1}{\Lambda^{d-4}} \sum^{n_d}_i
c^{(d)}_i(\mu) \, \hat{O}^{(d)}_i ,
\end{eqnarray}
where $\hat{O}^{(d)}_i$ denotes operators constructed from IR fields,
$\mathcal{L}_{d\leq4}$ are terms in the UV Lagrangian that remain after integrating out heavy fields,
$c^{(d)}_i$ denote effective (e.g.,\ Wilson) coefficients, $\Lambda$ is the UV energy scale, $\mu$
denotes the IR energy scale of interest, and $n_d$ is the finite number of operators at a given
dimension~\cite{Isidoristandard2024}. After averaging over heavy fields, the coefficients must be evolved
to the energy scale $\mu$ via renormalization group flow, which generically induces operator
mixing~\cite{DekensRenormalization2013,BhattacharyaDimension52015,BartocciRenormalisation2025}.
The result of this procedure is a dictionary that translates physics between energy
scales~\cite{deBlasEffective2018,GuedesEFT2024}. Particularly relevant for octupole-deformed nuclei, $\chi$EFT provides the bridge from hadronic CPV
operators to nuclear forces that generate Schiff moments. $\chi$EFT is built on the approximate chiral
symmetry of QCD, which is spontaneously broken by non-perturbative color confinement and explicitly
broken by quark masses and electroweak interactions. Below the chiral symmetry-breaking scale
$\Lambda_\chi\simeq1.2$~GeV, pions emerge as the relevant low-energy degrees of freedom and set the
power-counting scale. At energies below $\Lambda_\chi$,  chiral symmetry is both spontaneously broken by non-perturbative color confinement~\cite{AlkoferConfinement2009, deVriesIndirect2019} and  explicitly broken by the quark masses and electroweak interactions. Chiral symmetry breaking produces pions that set the IR energy scale for power counting. 

At energies above $\Lambda_\chi$, hadronic CPV is described by the quark–gluon Lagrangian. The
leading CPV terms, in addition to $\bar{\theta}$, arise from SMEFT sources including quark EDMs,
quark chromo-EDMs, the gluon chromo-EDM, and CPV four-quark interactions~\cite{Pospelov05,
EngelElectric2013,Chupp2019,DeVriesViolations2016,deVriesParity2020}. These Wilson coefficients are
run to low energies and matched onto $\chi$EFT LECs that parameterize the nuclear CPV potential $H_{CPV}$. Table~\ref{tab:uv_to_chipt} in the Supplemental material summarizes the mapping between SMEFT Wilson coefficients and the magnitude of the induced LECs.

Experimental data are insufficient to fit CPV LECs, which are therefore estimated via naive dimensional
analysis or computed using lattice QCD~\cite{CiriglianoRole2019,LiuLattice2025}. Recent progress in
lattice calculations of nucleon EDMs and CPV pion–nucleon couplings has significantly reduced
uncertainties, with extensions to additional CPV sources actively underway.

The final step in connecting $\chi$EFT to Schiff moments of octupole-deformed nuclei is provided by
nuclear structure calculations~\cite{EngelNuclear2025}. 
From eqs.~\ref{eq:schiff_pert} and \ref{eq:schiff_scaling}, the Schiff moment can be parameterized as~\cite{Dobaczewski2018, EngelNuclear2025}:
\begin{equation}
S_z=g_{\pi NN} (a_0 \bar{g}_0 + a_1 \bar{g}_1 + a_2 \bar{g}_2) + b_1 \bar{C}_1 + b_2 \bar{C}_2 + s_n\bar{d}_n + s_p\bar{d}_p,
\end{equation} \label{eq:schiff_pheno}
where $g_{\pi NN}\approx13$ is the strong $\pi NN$ coupling constant, and $a_i$, $b_i$, and $s_i$ are phenomenological coefficients encoding the results of nuclear structure calculations.

\begin{table*}[!tp]
\centering
\begin{threeparttable}
\footnotesize
\setlength{\tabcolsep}{2pt}
\renewcommand{\arraystretch}{1.05}
{
\begin{tabularx}{\textwidth}{Y Y Y}
\toprule
\textbf{SMEFT source} &
\textbf{Leading $\chi$EFT LECs} &
\textbf{Higher-order $\chi$EFT LECs} \\

\midrule

Strong CP $\bar\theta$ \xbreak
\quad Chiral-breaking \xbreak
\quad Isoscalar
& Pion-exchange \xbreak
\quad $\bar{g}_0 \simeq (15.5\pm2.6) \cdot 10^{-3}~\bar{\theta}$~\cite{deVriesParity2020} \xbreak 
\quad $\bar{g}_1 \simeq (3.4\pm2)\cdot 10^{-3} \bar\theta$~\cite{deVriesParity2020} \xbreak[1ex]
Short-range nucleon EDMs \xbreak
\quad $\bar d_0/\text{fm}\simeq 2\cdot 10^{-4}~\bar\theta$\tnote{a} \xbreak
\quad $\bar d_1/\text{fm} \simeq 6\cdot 10^{-4}~\bar{\theta}$\tnote{a} 
& 
Contact interactions \xbreak
\quad $\bar C_1/\text{fm}^3 \simeq -8\cdot 10^{-3} \bar \theta$~\cite{DeVriesViolations2016} \xbreak 
\quad $\bar C_2/\text{fm}^3 \simeq -1\cdot 10^{-3} \bar \theta$~\cite{DeVriesViolations2016} \xbreak[1ex]
Three-nucleon force \xbreak
\quad $\bar \Delta/\text{GeV} \sim \mathcal{O}(10^{-3}) \cdot \bar \theta$
\\
\midrule

Quark chromo-EDM $\tilde{d}_q$ \xbreak
\quad Chiral-breaking \xbreak
\quad Isoscalar $\tilde{d}_0 \approx \frac{1}{2}\left(\tilde{d}_u+\tilde{d}_d \right)$ \xbreak
\quad Isovector $\tilde{d}_1 \approx \frac{1}{2}\left(\tilde{d}_u-\tilde{d}_d \right)$
& Pion-exchange \xbreak
\quad $\bar g_0 \simeq \begin{cases}
    [-94,55]$~\cite{Bhattacharyadetermination2025}$ \\
    [-11,28]$~\cite{PospelovBest2002}$
\end{cases} \hspace{-0.75em} \cdot~\tilde d_0/\text{fm} $\xbreak
\quad $\bar g_1 \simeq \begin{cases}
    [-41,-530]$~\cite{Bhattacharyadetermination2025}$ \\
    [22,120]$~\cite{PospelovBest2002}$
\end{cases} \hspace{-0.75em} \cdot~\tilde d_1/\text{fm}$ \xbreak[1ex]
Short-range nucleon EDMs \xbreak
\quad $\bar d_{0,1} \sim \mathcal{O}(10^{-1})\cdot e \tilde d_{0,1}$
& Contact interactions \xbreak
\quad $\bar C_{1,2}/\text{fm}^3 \sim \mathcal{O}( 10^{-2})\cdot \tilde d_{0,1}/\text{fm}$
\xbreak[1ex]
Three-nucleon force \xbreak
\quad $\bar \Delta/\text{GeV} \sim \mathcal{O}(10^{-4}) \cdot \tilde d_{0,1}/\text{fm}$

\\
\midrule

Quark EDM $d_q$ \xbreak
\quad Chiral-breaking \xbreak
\quad Isoscalar $d_0\approx\frac{1}{2}\left(d_u+d_d\right) $\xbreak
\quad Isovector $d_1\approx\frac{1}{2}\left(d_u- d_d\right) $
& Short-range nucleon EDMs \xbreak
\quad $\bar d_0 \simeq 1.2 \cdot d_0$~\tnote{b} \xbreak
\quad $\bar d_1 \simeq 2.0 \cdot d_1$~\tnote{b}
&  Pion-exchange \xbreak
\quad $\bar g_{0,1} \sim \mathcal{O}(10^{-2}) \cdot d_{0,1}/e\text{fm}$ \xbreak[1ex]
Contact interactions \xbreak
\quad $\bar C_{1,2}/\text{fm}^3 \sim \mathcal{O}( 10^{-5})\ d_{0,1}/e\text{fm}$
\xbreak[1ex]

Three-nucleon force \xbreak
\quad $\bar \Delta/\text{GeV} \sim \mathcal{O}(10^{-7}) \cdot d_{0,1}/e\text{fm}$
\\
\midrule

Four-quark left-right 
$\Xi_1 \simeq 0.8 \cdot \Xi_8$\tnote{c}
\xbreak
\quad Chiral-breaking \xbreak
\quad Isovector
& Three-nucleon force \xbreak
\quad $\bar \Delta /\text{GeV} \simeq (2.3\pm0.1)\cdot 10^{-5} \cdot 
\Xi_1$\cite{deVriesParity2020} \xbreak[1ex]

Pion-exchange \xbreak
\quad $\bar g_0 \sim \mathcal{O}(10^{-8}) \cdot \Xi_1$ \xbreak
\quad $\bar g_1 \sim \mathcal{O}(10^{-6})\cdot \Xi_1$ 
& Short-range nucleon EDMs \xbreak
\quad $\bar d_{0,1}/\text{fm} \sim \mathcal{O}(10^{-8}) \Xi_1$
\xbreak[1ex]
Contact interactions \xbreak
\quad $\bar C_{1,2}/\text{fm}^3 \sim \mathcal{O}(10^{-9}) \cdot \Xi_1$

\\
\midrule

Gluon chromo-EDM $\tilde d_G$ \xbreak 
Four-quark pseudoscalar-scalar $\Sigma_{1,8}$ \xbreak
\quad Chiral-invariant \xbreak
\quad Isoscalar
& Contact interactions \xbreak
\quad $\bar C_{1,2}/\text{fm}^3 \sim \mathcal{O}(1) \cdot \tilde d_G/\text{fm}^2 $ \xbreak
\quad $\bar C_{1,2}/\text{fm}^3 \sim \mathcal{O}(10^{-7}) \cdot \Sigma_{1,8} $ \xbreak[1ex]
Short-range nucleon EDMs \xbreak
\quad $\bar d_{0,1}/e\text{fm} \sim \mathcal{O}(10^{-1}) \cdot \tilde d_G/\text{fm}^2$ \xbreak
\quad $\bar d_{0,1}/e\text{fm} \sim \mathcal{O}(10^{-8}) \cdot \Sigma_{1,8}$
& Pion-exchange \xbreak
\quad $\bar g_{0,1} \sim \mathcal{O}(10^{-1}) \cdot \tilde d_G/\text{fm}^2$\xbreak
\quad $\bar g_{0,1} \sim \mathcal{O}(10^{-8}) \cdot \Sigma_{1,8}$
\xbreak[1ex]
Three-nucleon force\xbreak
\quad $\bar \Delta/\text{GeV} \sim \mathcal{O}(10^{-3}) \cdot \tilde d_G/\text{fm}^2$ \xbreak
\quad $\bar \Delta/\text{GeV} \sim \mathcal{O}(10^{-8}) \cdot \Sigma_{1,8}$

\\
\bottomrule
\end{tabularx}
}
\caption{\textbf{Matching SMEFT Wilson coefficients for CPV to $\chi$EFT LECs and their magnitudes.} Each source is classified according to its chiral ($L \leftrightarrow R$) and isospin ($u \leftrightarrow d$) symmetries, which affect the resultant LEC hierarchy. Relations given by $\sim$ are obtained using Naive Dimensional Analysis (NDA) from Refs.~\cite{deVrieseffective2013, deVriesParity2020}. Two ranges are provided for $\bar g_{0,1}(\tilde d_{0,1})$ reflecting recent work on a previously neglected form factor~\cite{Bhattacharyadetermination2025}, with a different sign convention for $\bar g_1$. The Wilson coefficients $\tilde d_G$ and $\Sigma_{1,8}$ are grouped together as they cannot be distinguished from LEC hierarchy measurements alone.}
\begin{tablenotes}
    \item[a] Obtained using the NLO formula for $d_{n,p}$~\cite{deVriesParity2020}, and calculated lattice QCD values for $\bar g_{0,1} (\bar \theta)$~\cite{deVriesParity2020} and $d_{n,p}(\bar \theta)$~\cite{LiuLattice2025}. 
    \item[b] Obtained by setting $\bar{g}_{0,1}=0$ and using the tensor charge calculation of $d_{n,p}$ from Ref.~\cite{deVriesParity2020}.
    \item[c] $\Xi_{1}$ and $\Xi_{8}$ are not linearly independent, being induced at low energies after integrating out $W$ bosons~\cite{DekensRenormalization2013}
\end{tablenotes}
\label{tab:uv_to_chipt}
\end{threeparttable}
\end{table*}
\end{document}